\documentclass[9pt,shortpaper,twoside,web]{ieeecolor}
\usepackage{generic}
\usepackage{amsmath,amssymb,amsfonts}
\usepackage{algorithmic}
\usepackage{graphicx}
\usepackage{textcomp}
\usepackage{xspace}
\usepackage{float}
\usepackage{tabularx}
\usepackage{booktabs}
\usepackage{lipsum}
\usepackage{hyperref}
\usepackage{setspace}

\newcommand{\tabRef}[1]{Table \ref{#1}\xspace}
\newcommand{\figRef}[1]{Fig. \ref{#1}\xspace}

\newcommand{\ESA}[0]{Evaluation-Subset-A\xspace}
\newcommand{\ESB}[0]{Evaluation-Subset-B\xspace}

\usepackage[labelsep=period]{caption}
\renewcommand{\thetable}{\Roman{table}}

\usepackage[backend=biber,bibencoding=utf8,style=ieee]{biblatex}
\bibliography{bibliography/references.bib}

\newcolumntype{Y}{>{\centering\arraybackslash}X}

\usepackage{amsmath,amssymb,amsfonts}
\usepackage{algorithmic}
\usepackage{graphicx}
\usepackage{textcomp}
\def\BibTeX{{\rm B\kern-.05em{\sc i\kern-.025em b}\kern-.08em
    T\kern-.1667em\lower.7ex\hbox{E}\kern-.125emX}}
\usepackage[utf8]{inputenc}
\usepackage[T1]{fontenc}

\begin{document}
% \history{Date of publication xxxx 00, 0000, date of current version xxxx 00, 0000.}
% \doi{10.1109/ACCESS.2017.DOI}
\title{Automatic Non-Invasive Isolation of Respiratory Cycles}
\author{%
\uppercase{Benedikt Holm},
\uppercase{María Óskarsdóttir},
\uppercase{Erna S. Arnardóttir},
\uppercase{Marta Serwatko},
\uppercase{Jacky Mallett}, 
and \uppercase{Michal Borsky}%
\thanks{This work was supported by Icelandic Research Fund \#175256-0611), Nordforsk (NordSleep
project 90458) via The Icelandic Research Fund and the European Union’s Horizon 2020 research and innovation programme
under grant agreement no. 965417,}
\thanks{B. Holm is with Reykjavik University, School of Technology, Department of Computer Science, Reykjavik, Iceland.  (e-mail: b@spock.is).}
\thanks{M. Óskarsdóttir is with Reykjavik University, School of Technology, Department of Computer Science, Reykjavik, Iceland.}
\thanks{E. S. Arnardóttir is with Reykjavik University, School of Technology, Reykjavik University Sleep Institute \& Landspitali - The National University Hospital of Iceland, Reykjavik, Iceland.}
\thanks{M. Serwatko is with Reykjavik University, School of Technology, Reykjavik University Sleep Institute, Reykjavik, Iceland.}
\thanks{J. Mallett is with Reykjavik University, School of Technology, Department of Computer Science, Reykjavik, Iceland.}
\thanks{M. Borsky is with Reykjavik University, School of Technology, Department of Computer Science, Reykjavik, Iceland.}
}

\maketitle
\begin{abstract}
In this paper, we introduce a novel algorithm designed to isolate individual respiratory cycles on a thoracic respiratory inductance plethysmography signal. The algorithm locates breaths using signal processing and statistical methods and enables the analysis of sleep data on an individual breath level. The algorithm was evaluated on 7.3 hours of hand-annotated data, or 8782 individual breaths in total, and was estimated to correctly isolate 94\% of respiratory cycles while producing false positives that amount to only 5\% of the total number of detections. The algorithm was specifically evaluated on data containing a great number of sleep-disordered breathing events. We found that the algorithm did not suffer in terms of accuracy when detecting breaths in the presence of sleep-disordered breathing.
The algorithm was also evaluated across a large set of participants, and we found that the accuracy of the algorithm was consistent across participants. This algorithm is finally made public via an open-source Python library. 

\end{abstract}

\begin{IEEEkeywords}
% Enter key words or phrases in alphabetical 
% order, separated by commas. For a list of suggested keywords, send a blank 
% e-mail to keywords@ieee.org or visit \underline
% {http://www.ieee.org/organizations/pubs/ani\_prod/keywrd98.txt}
Computer aided analysis, signal processing, polysomnogram analysis, breath segmentation.
\end{IEEEkeywords}

\section*{Introduction}

\PARstart{W}{hen} patients are suspected to suffer from a sleep disorder, they are recommended to undergo polysomnography~(PSG), which is an overnight collection of various physiological signals, such as respiratory inductance plethysmography~(RIP), oxygen saturation~(SpO2), nasal airflow, electroencephalography (EEG), electromyograms (EMG), electrocardiography (ECG), audio, and others~\cite{aasm2014}.
This type of study is performed either in the controlled environment of a hospital, or in a home setting, each with their advantages and disadvantages \cite{doi:10.1164/ajrccm.162.3.9908002}. 
To be able to correctly detect a sleep disorder, an expert sleep scorer must manually review the PSG and annotate sleep stages and events of interest. These include respiratory events (apneas, hypopneas), oxygen desaturations, movements, and respiratory event related arousals. 
Finally, the annotations are used to determine the diagnosis and recommend a treatment. 
This manual annotation is tedious, and not all scorers will agree on a given annotation.  The average agreement for sleep stages is, for example, only 82.6\%\cite{rosenberg_vanhout_2013}. There is therefore a growing need for automation and large-scale analysis in the field of sleep medicine \cite{ARNARDOTTIR2021447}.

The field of computerised scoring of PSG data mainly revolves around the analysis of fixed-length epochs, the alternative being called adaptive segmentation \cite{Praetorius1977-sl}. The main bulk of research into adaptive segmentation on polysomnography data has been in the context of examining brain activity during sleep \cite{Schulz2008-jm}, and has for example, seen some success in sleep staging \cite{Prochazka2017-td, Koch2019-rp}.

\begin{figure}[H]
    \centering
    \includegraphics[scale=0.5]{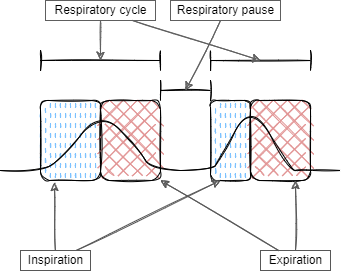}
    \caption{Phases of the respiratory cycle.}
    \label{fig:breath-phases}
\end{figure}

The common definition of a respiratory cycle in the literature splits a single cycle into 4 distinct phases; inspiratory, inspiratory pause, expiratory, and expiratory pause \cite{Hult2000-lx}. 
In this work, a single cycle in the respiratory system is defined as starting with an inhalation, and ending just after the following exhalation.
The terms 'breath' and 'respiratory cycle' are considered synonyms in this work. 
This definition ignores the inspiratory pause phase, and interprets the expiratory pause phase rather as a pause between two individual breaths belonging to neither.
Furthermore, this definition dictates that no two breaths can occupy the same moment in time. The different phases of the respiratory cycle as defined in this work are visualised in \figRef{fig:breath-phases}.

This work presents and evaluates a novel algorithm designed to locate individual respiratory cycles within a commonly collected signal in a PSG. 
This paper defines the term for this task as respiratory cycle isolation~(RCI).

The existing literature on performing RCI is sparse, and very few papers exist that describe algorithms capable of performing the task.
Moyles \& Erlandson~ proposed a non-parametric statistical approach to RCI, detecting changes in the trend of the air flow signal \cite{95756}. They did not perform any validation of their algorithm.
% 
% Chervin et al. described a custom breath cycle detection algorithm based on calculating local maxima, minima and endpoints on the air flow signal \cite{Chervin2004-at}. 
% 
Lopez-Meyer et al. presented a RCI algorithm based on peak and valley detection in the RIP signals to determine the beginning and end of a breath segment \cite{LopezMeyer_2011}. They reported a 96\% precision in detecting breath cycles for participants during rest.
Rosenwein et al. introduced a breath detection algorithm based on a random forest approach \cite{Rosenwein2014-nz}. They derived 351 features from audio recordings and trained the model to detect inspirations and exhalations, reporting an 87\% and  76\% accuracy in predicting inspiration and expiration events, respectively. 
A Python library for RIP belt analysis and RCI exists, designed for use on people engaged in conversation, which also uses a peak and valley location algorithm to find respiratory cycles \cite{wlodarczak2019respinpeace}. The paper describing the algorithm makes no reference to validation.
Palaniappan et al.  developed a neural network solution for classifying respiratory phases using respiratory sounds \cite{Palaniappan2017-jv}. They reported very good overall performance for their model, but did not disclose the accuracy of the model in terms of the different phases. 
Finally, Hult et al. proposed a bioacoustic method that can accurately time respiration from tracheal sounds, using a summation method over the frequency domain of the audio signal. 
They evaluate their algorithm on 2074 respiratory cycles from two groups of participants, one being recorded in a quiet environment and another with acoustic disturbances from surrounding activities. They report detecting respiratory phases with 99\% accuracy on participants in the group with the quieter environment, and approximately 90\% accuracy in the group recorded with more noisy conditions.
% "with an accuracy that is greater than 88\%", but also report results as high as 99\% \cite{Hult2004-wb}. 

Although the existing literature presents diverse methods for RCI algorithms, research in the field is quite limited, and the optimal solution is not known, 
% { fiffa i thessu
% matthildur er cutie pie
other than a clear need for a focus on validation. 
%} 
A noticeable trend in the articles cited above is a lack of clarity regarding the verification and validation of the RCI algorithm they present. Some articles provide only a vague explanation how the breath placement accuracy was calculated, and some rely solely on visual observation of the output rather than objective verification. Few articles present the performance of their algorithms on data that features events such as apnea, snoring, or movement.

The existing methods of RCI are mainly based on statistics, peak and valley detection in the airflow, thoracic or abdominal RIP signals, or feature extraction and modelling, mainly done on the audio signal \cite{95756,Chervin2004-at,LopezMeyer_2011,Rosenwein2014-nz,wlodarczak2019respinpeace,Hult2000-lx, Hult2000-lx}.
In literature, the task of RCI is sometimes referred to as breath segmentation \cite{LopezMeyer_2011,  orarson2021-js}, or breath cycle segmentation \cite{Palaniappan2017-jv}, however this term leaves ambiguity on whether the segmentation is done on a respiratory phase basis, or on a respiratory cycle basis. 
For that reason, this paper terms the task as RCI, rather than breath segmentation.
% For that reason, the term used for the 

% 

As previously stated, the PSG contains a multitude of signals, but in the context of this paper, however, only the airflow and RIP belt signals are relevant. The airflow signal measures nasal respiration, and is most commonly measured with a pressure transducer attached to a nasal cannula, and is used to detect respiratory events \cite{aasm2014}. 
The RIP  signals are measured via two belts that stretch around the thorax and abdomen to measure the change in inductance caused by the movement of the body part they are placed around. RIP belts are used to detect respiratory events in conjunction with the nasal cannula and estimate respiratory effort\cite{aasm2014}.
Although not guaranteed to be included in a PSG, the audio signal is measured with a microphone, and is commonly used for snore monitoring \cite{aasm2014}.
% 
% Less common in a PSG is the thermistor signal, which measures the 

The main contributions this paper seeks to present are:
\begin{itemize}
    \item Introduce a novel algorithm for performing RCI.
    
    \item Propose new methods for evaluating the performance of algorithms designed to locate events in signals. 
    
    \item Present the results from  the introduced algorithm using the proposed evaluation methods.
\end{itemize}

\section*{Methodology}
    
To perform RCI, a decision must be made on which signal is most appropriate to perform the task on. The two main factors in the decision are the failure rate of the signals and any potential effects of external factors.
The airflow signal was deemed inappropriate for this work because the nasal cannula has several logistical issues. The sensor can get loose, affecting the measured airflow, or the participant can start mouth breathing , bypassing the sensor completely. 
% The same issues apply to the thermistor signal, which measures the heat generated by expiration from the mouth, the sensor can become dislodged, or the patient can start breathing with their nose. 
Multiple studies show that the nasal airflow signal exhibits poor quality in 10\% of cases \cite{Portier2000-iy, BaHammam2005-dv}. 
The audio signal was also eliminated, even though some studies show the signal is not as prone to error as the other signals \cite{BaHammam2005-dv}, because the signal may contain a plethora of different acoustic events such as snoring, movement-related artefacts, or various background noises which complicate the task \cite{Hult2004-wb}.  
Using RIP belts has the advantage that they are not susceptible to ambient noises, nor the bypass problem that the airflow signal may encounter.
Of the two RIP belts, the thoracic RIP signal captures the action of the chest-wall muscles more closely than the abdominal signal. 
For the reasons listed above, the thoracic RIP signal was chosen as the most appropriate signal to perform RCI on.
    % This section is divided into three parts; The overall design of the algorithm, the evaluation data used got the algorithm, and the methods used to evaluate the algorithm.
    
\subsection*{Evaluation Data}
    % Since the variety of different situations the algorithm might be expected to have to deal with during sleep, a
    An extensive evaluation of the correctness of the algorithms output is required, both during normal breathing, and other conditions that may arise during sleep. 
    The dataset used for validation contained 31 overnight PSGs from a population of people diagnosed with obstructive sleep apnea, as well as people with no known sleep issues (VSN-14-080). 
    Of the participants, 13 were female and 18 were male. 
    The mean age of the participants was 47.1 years, in the range of 20 - 69 years.
    The mean body-mass index~(BMI) was 29.9 kg/$\text{m}^{2}$ in the range of 21.6 - 49.3 kg/$\text{m}^{2}$. 
    The mean apnea-hypopnea index~(AHI) was 9.3$h^{-1}$ in the range of 0.0 to 34.8 $h^{-1}$.  
    Due either to signal failure in the RIP signal, or errors in exporting the recordings from the proprietary NOX format to the standard European data format~(EDF), 5 recordings had to be discarded.
    Each PSG included all standard signals, including EEG, EOG, EMG, ECG, airflow recorded with a nasal cannula, thorax and abdomen RIP belts, pulse oxymetry~(SpO2), and an audio signal.
    The RIP belts in the dataset were recorded with a 25 $\mathit{Hz}$ sampling frequency. 
    Additionally, esophageal pressure was recorded with a nose fed catheter \cite{serwatko2016validation}.

    The algorithm was evaluated against 39 variable-length manually annotated evaluation intervals, which further split into 2 evaluation subsets. 
    The first set, referred to as \ESA
    was selected to specifically contain various sleep disordered breathing~(SDB) events, as well as different sleep stages. \ESA contained 14 variable length intervals with the mean length of 16 minutes, in the range of 1.5 to 37.5 minutes, with the cumulative length of 225.65 seconds (3.6 hours). The SDB events in \ESA included obstructive apneas, hypopneas, and increases in respiratory effort without apnea or hypopnea. Further events included in \ESA were sleep stages, movements, oxygen desaturations and snoring.
    
    % ESA CUT {
    % The 14 intervals that make up \ESA are listed in Table \ref{tab:evaluation_periods}.
    % \input{Tables/evaluation_subset_A}
    % }
    
    These intervals, however, were only selected from one participant in the dataset, and are not representative of the general public. 
    To address that issue, a second evaluation subset was defined, referred to as \ESB, consisting of a collection of 10 minute intervals from the remaining 25 valid PSGs in the dataset. 
    These intervals were selected randomly from each recording in order to avoid cherry-picking favourable intervals. 
    The random selection was restricted to selecting from 1 hour after the recording starts to 1 hour before the recording ends. This was done to reduce the probability of including either the participant settling down to sleep, or moving around as they wake up. 
    
    % The placement of the intervals from \ESA over night is visualised in figure \figRef{fig:ESB_distribution}.
    % \begin{figure}[H]
    %     \centering
    %     \includegraphics[width=\linewidth]{Graphics/periods.png}
    %     \caption{Distribution of \ESB intervals}
    %     \label{fig:ESB_distribution}
    % \end{figure}
    
    The locations of individual breaths in both evaluation subsets were then manually marked using a custom made scoring tool programmed in Python. The manual breath annotations represented the ground truth.
    Of the total 39 intervals in \ESA and \ESB, one was found to be incorrectly manually annotated and was discarded. The algorithm was therefore evaluated on 7.3 hours of manually annotated data over 38 intervals, containing 8782 individual breaths from 26 participants.
    
    \subsection*{Algorithm Design}
\begin{figure}[h]
        \centering
        \includegraphics[width=0.95\linewidth]{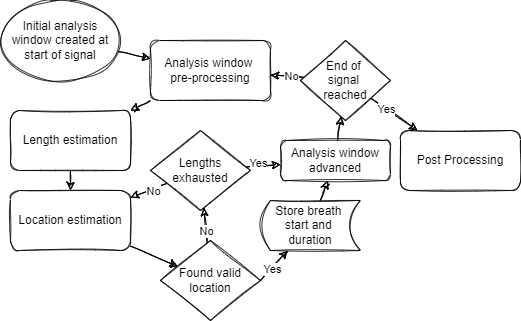}
        \caption{Respiratory Cycle Isolation algorithm flowchart.}
        \label{fig:BSS flowchart}
    \end{figure}

    A flowchart of the proposed RCI algorithm is presented in Figure \ref{fig:BSS flowchart}.
    % , and its pseudo code is presented at the end of this section. 
    The algorithm takes a thoracic RIP signal as a parameter, along with a sampling frequency $\mathit{fs}$. The output is a list of individual respiratory cycles, each consisting of the onset, i.e. the start of the respiratory cycle in seconds since the signal start, and the duration of the respiratory cycle in seconds. 
    The algorithm works on the principle of segmenting the signal into short windows $w[n]$, with arbitrary onset $n$ in the signal, and searching for a single respiratory cycle in the selected window.
    The algorithm first calculates the autocorrelation function for $w[n]$ in order to estimate the lengths $\textit{l}$ of all potential breaths in the window.
    It then uses a probability model to discard breath length candidates that are considered too unlikely, either because the length is too long or too short. 
    Then, for each remaining breath length $\textit{l}$, the algorithm creates a template waveform of that length, which it correlates with the signal window to find where in the window the breath onset is most likely to be.
    After the window is analysed, the algorithm advances the window further in the signal, and repeats the process. The analysis windows overlap in order to allow the algorithm to analyse every breath multiple times.
    
    % Before the algorithm can start locating breaths within a given RIP signal, it requires two main parameters to be defined; 1) the analysis window length, and 2) window overlap.
    % % 
    % In order to determine the optimal window size, the manual labels in \ESA and \ESB were used to determine that the mean breath length was 3.53 seconds. 
    % % 
    % Therefore, 8 seconds was chosen as the window length, which was slightly more than twice the mean breath length. This specific length was chosen in order to increase the probability that any $w[n]$ contains at least two breath cycles. The overlap, which determines how much of the previous window is included in the next window, was chosen to be 95\% via experimentation.
    % % { 
    % The Shannon-Nyquist sampling theorem dictates that for a continuous signal to be discretely reconstructed without error, the sampling frequency must be twice as large as the signal's maximum frequency \cite{shannon_communication}. 
    % % 
    % The RIP signal in the dataset was sampled with 25 $Hz$ sampling rate, and the average period of a breath cycle is 3.53 seconds. The frequency $F$ of a signal with periodicity $T$ can be calculated: 
    % \begin{equation}
    %     F = \frac{1}{T}
    % \end{equation}
    % The average frequency of respiration captured by the RIP signal is therefore approximately 0.283 $Hz$, far less than the 12.5 $Hz$ maximum frequency imposed by the Shannon-Nyquist theorem. Additionally, in order to violate this constraint, a breath must be less than 0.08 seconds long, which was never observed in the data.
    
     % }

    \subsubsection*{Signal Pre-processing}
    
    % The periodicity $\textit{T}$ of the RIP signal is inversely related to the rate of breathing, and is therefore an estimation of the length of a possible target breath cycle within the $w[n]$. 
    % To ensure that its estimation was as accurate as possible, it was vital to prevent the autocorrelation from creating artificial peaks due to the presence of artifacts in the signal.
    % The pre-processing of the signal achieve this, 
    The preprocessing of the signal was two-fold, first the entire signal is smoothed, and then skew was removed from the signal.
    The RIP signal was smoothed using a Savitzky-Golay filter, which fits a polynomial function to smooth the data points~\cite{Savitzky1964-zo}.
    The filter employed used a third order polynomial and a window size of 51. 
    The filter parameters were tuned beforehand via experimentation to ensure that the smoothing had a minimal effect on the overall shape of the RIP signal while still eliminating some of the finer noise. 
    Additionally, each individual $w[n]$ was corrected for skew.
    This was achieved by fitting a linear function to the signal in $w[n]$, then adjusting the function such that the y-intercept was 0.0. 
    The function was then subtracted from each sample of the signal. 
    This procedure removes large-scale skew from the signal window, but leaves the general shape of the signal intact.
    The procedure also had a positive effect on the template waveform fitting procedure, making it less likely to produce incorrect results due to skew.
    The result of this pre-processing step was that the cleaned thoracic RIP signal was ready to be used to estimate breath lengths. 
    % 
    % Figure \ref{fig:smoothing} shows a thorax signal before and after the skew removal.
    
    % \begin{figure}[h!]
    %     \centering
    %     \includegraphics[width=0.35\linewidth]{Graphics/smoothing_operation.png}
    %     \caption{Thoracic RIP before and after skew removal.}
    %     \label{fig:smoothing}
    % \end{figure}
    
    \subsubsection*{Main Algorithm Body}
    In the first step, the algorithm takes an analysis window $w[n]$, containing a cleaned signal and
    estimated its periodicity $T$ using the autocorrelation function.
    % , which is commonly used to estimate a signal's periodicity \cite{doi:10.1137/1.9781611972757.40}.
    % 
    The principle of the autocorrelation function~(ACF) is to shift the signal forwards in time by $k$ and to compare it to itself. 
    When $k = 0$, the signal correlates perfectly with itself, but as $k$ increases, the correlation decreases. 
    The formula for autocorrelation of a signal $x$ is:
    \begin{equation} \label{eq:autocorrelation}
        \text{ACF}(x)[k] = \frac{1}{N} \sum_{n=0}^{N-1} x[k] \cdot x[n+k],
    \end{equation}
    \noindent
    where N is the length of $x$, and $k$ the shift.
    For periodic signals, when $k = \frac{T}{2}$, the value of the auto correlation is low, as the signal is being compared to itself when it is in asynchony.
    As $k$ approaches $T$, the correlation value increases, as the first period lines up with the following period.
    Thus the peaks of $ACF(w[n])$ can be used to estimate the periodicity of a signal \cite{doi:10.1137/1.9781611972757.40}. 
    In the next step, the peaks in $ACF(w[n])$ were found using a peak-finding algorithm. Since the analysis window length was more than twice the mean breath length, $w[n]$ is likely to contain at least two breath cycles, and thus multiple peaks.
    % In that case, it is probable that $ACF(w[n])$ has more than one peak as can be seen in Figure \ref{fig:peaks_in_autocorrelation}.
    % 
    % \begin{figure}[H]
    %     \centering
    %     \includegraphics[scale=0.6]{graphics/figures/part_1/bss_algorithm_graphics/rip_autocorrelation.png}
    %     \caption{Illustration of one-sided autocorrelation of RIP signal window.}
    %     \label{fig:peaks_in_autocorrelation}
    %     % \textit{Peaks as found by scipy's find\_peaks method are marked on the autocorrelation}
    % \end{figure}
    % 
    To address the possibility of false alarm peaks produced by this approach, the algorithm models the breath length probabilities with a normal distribution.
    The parameters $\mu$ and $\sigma$ for the normal distribution were calculated as the mean and standard deviation of the length of breaths within \ESA and B.
    % 
    % were estimated using the lengths of \ESA and \ESB.
    % 
    % {\Huge write this out clearer, sigma is what.... These values are calculateds as the standerd deviation and the mean of the breath lengths in A and B}
    % 
    % To estimate the probability of a particular breath of length $l$, the following function can be used:
    % 
    % To estimate the likelihood of a breath length $l$, \eqref{eq:pdf} can be used:
    
\begin{figure}[H]
        \centering
        \includegraphics[width=0.80\linewidth]{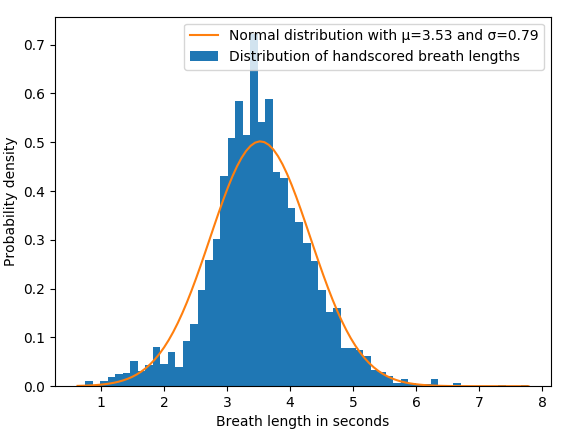}
        \caption{Reference breath length histogram with model normal distribution.}
        \label{fig:breath_length_pdf}
    \end{figure}

    % \begin{equation} \label{eq:pdf}
    %     \text{p(l)} = N(l|\mu,\sigma) = \frac{1}{0.79\sqrt{2\pi}}e^{-\frac{1}{2}(\frac{l-3.53}{0.79})^{2}}.
    % \end{equation}
    % 
    % \noindent
    The modelled probability distribution can be seen in Figure \ref{fig:breath_length_pdf} along with a density histogram of the breath lengths from the sets of 8782 manually annotated intervals.
    % As the figure shows, the modelled distribution is an approximation, but upon visual inspection was considered to be good enough for the purposes of this algorithm.

    Using this probability distribution, the algorithm can rank the breath length candidates, ensuring that it considers the most probable breath length first, thus saving on computing time. 
    Additionally, any breath length candidate whose length probability is less than three standard deviations from the mean is discarded as being too unlikely. 
    % The value for this elimination criterion was found via experimentation. 
    Practically speaking, this means that any breath shorter than approximately 1 second or longer than 6 seconds was discarded. 
    % 
    % In the case where a breath candidate is discarded because it is too long, the most likely reason was that it represents a second breath in the signal window.
    % An example of this is peak 2 in Figure \ref{fig:peaks_in_autocorrelation}.
    % This filtering was not found to have detrimental effects on subsequent breath length estimations, since the windows overlap considerably, so the algorithm may have multiple attempts to detect any missed breath when it advances to the next analysis window.
    %

    \noindent    
    In the next step, for every remaining breath length candidate, a discrete sine template waveform is generated, using the following formula:
    
    \begin{equation}
        sin[n] = sin( \frac{n * 2 * \pi}{l} +  \theta ), 
    \end{equation}
    \noindent
    where $l$ is the length of the candidate in seconds, and $\theta$ is an offset that can be set to 1.5 $\pi$ to shift the waveform so that it starts at -1, ends at -1, and has a peak in the middle.  
    To find where a given template waveform fits most closely to the RIP signal window, the algorithm compares it to the RIP signal using the Pearson~correlation~coefficient~($\rho$) at each point on the RIP signal. The formula for calculating $\rho$ for a pair of signals is:

    \begin{equation} \label{eq:PCC}
        \rho \text{(w[n], sin[l])} = 
        \frac{\text{cov(w[n], sin[l])}}{ \sigma_{w[n]}\sigma_{sin[l]} }
    \end{equation} 
    where cov(w[n],sin[l]) is the covariance of the window and template waveform , and the covariance can be calculated as:
    \begin{equation} \label{eq:PCC_COV}
        \text{cov(w[n], sin[l])} = 
        \sum_{i=1}^{N} \frac{(w[n][i] - \bar{w[n]})(sin[l][i] - \bar{sin[l]})}{N}
    \end{equation}
    where N is the length of w[n] and sin[l] which must be equal, and $\bar{w[n]}$ and $\bar{sin}$ are the means of the respective signals.
    The sign of $\rho$ describes whether the signals are positively or negatively correlated, and its value describes how strong the correlation is. A $\rho$ value of  0.0 means that the signals are not correlated, a value of 1.0 indicates a positive correlation, and a value of -1.0 means that the variables are perfectly inversely correlated. The algorithm treats the RIP signal as one variable and the template waveform signal as another, and calculates the correlation of the template waveform over the entire window.
    % This method is illustrated in Figure \ref{fig:breath_approximation}. 
    The correlation of the template waveform and the RIP signal produces a third signal, whose peaks represent possible onsets of the target breath.
    
    % \begin{figure}[h!]
    %     \centering
    %     % \includegraphics[scale=0.95]{graphics/figures/part_1/breath_approximation_acf.png}
    % \includegraphics[scale=0.55]{graphics/figures/part_1/bss_algorithm_graphics/breath_placement.png}
    %     \caption{Breath template, thoracic RIP signal, and the correlation of the template waveform with the signal.}
    %     \label{fig:breath_approximation}
    % \end{figure}

    Since the template waveform is an approximation of the shape of a breath, $\rho$ is not expected to reach 1.0. However, the correlation still provides information about the validity of the breath onset. 
    The algorithm discards any breath onset candidate whose $\rho$ is less than 0.75. The $\rho$ elimination criterion  was chosen via experimentation to eliminate as many inaccurate guesses as possible, while still not being so strict as to eliminate legitimate guesses on noisy data.
    If this elimination step filters out all breath onset candidates, the algorithm repeats the template waveform fitting process with another breath length candidate.
    If the algorithm processes all breath length candidates and no breath is found in the current signal window, then the algorithm moves on to the next window.
    If the correlation is above the threshold, the algorithm adds the onset and the duration to a list of breaths and moves the window onset to the end of the detected breath. This process is repeated until the signal is fully analysed.

\subsubsection*{Breath Placement Post-processing} \label{sec_p1:methods_post_processing}

\begin{figure}[h]
        \centering
        \includegraphics[width=0.6\linewidth]{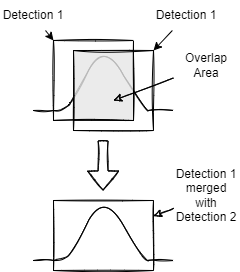}
        \caption{Detection merging procedure}
        \label{fig:prediction_merge}
    \end{figure}

    As the sliding windows overlap, the algorithm has a tendency to re-discover breaths. To solve this problem, the $i^{th}$ breath is compared to the $i+1^{th}$ breath. If the overlap of the breaths exceeds 80\% of the total length, the breaths are considered a double detection, and therefore the detections are merged. The process of merging two breath detections involves replacing them with a single detection which covers the area that both previous detections covered.
    The percentage overlap calculation for a pair of time spans is:
    \begin{equation}
        O_{w}(A, B) = \frac{2|A {\displaystyle \cap} B|}{|A|+|B|} 
        \label{eq:overlap_percentage}
    \end{equation}
    % overlap = 
    % \begin{equation}
    % max(0, 2 * \frac{
    %     min(\text{e}_{i}, \text{e}_{i+1}) - max(\text{s}_{i}, \text{s}_{i+1})
    %     }
    % {\text{dur}_{i} + \text{dur}_{i+1}}),
    % \label{eq:overlap_percentage}
    % \end{equation}
    % 
    \noindent
    % where s$_{i}$, e$_{i}$ and dur$_{i}$ are the start, end, and duration of the $i^{th}$ detection.
    where $|A|$ and $|B|$ are the lengths of time spans $A$ and $B$ respectively, and $|A {\displaystyle \cap} B|$ is the overlap area of detections $A$ and $B$
    If the breaths do not overlap at all, the value produced by this function is negative, and in the case of perfect overlap, the overlap value is 1.0. 
    For this reason, the function is clamped above 0.0. 
    The detection merging procedure is visualised in \figRef{fig:prediction_merge}.

\begin{figure}[h]
        \centering
        \includegraphics[width=0.7\linewidth,keepaspectratio]{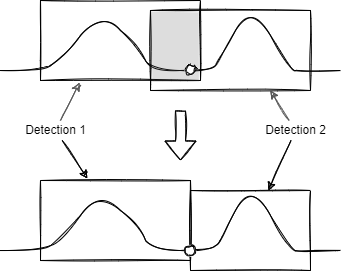}
        \caption{Overlap elimination process (minimum of the RIP signal marked by a white circle)}
        \label{fig:overlap_elimination}
    \end{figure}

    Due to the 80\% overlap required to merge breaths, the breaths can still overlap by up to 20\%. 
    By definition, a breath cannot overlap with another breath, so for each breath, the $i^{th}$ breath is compared to the $i+1^{th}$ breath.
    If they still overlap, the end of the $i^{th}$ and start of the $i+1^{th}$ breath are moved to the time of the minimum value of the RIP signal within the overlapping region. This process is visualised in \figRef{fig:overlap_elimination}.
    
    The end result of this post processing process is that there is no overlapping pair of detections, satisfying the constraint that no two breaths can share a moment in time. 
    When run on \ESA and \ESB, the post processing step removed 2.05\% of detections on average from each interval.
    
    % The process of creating the manual labels was
    % \clearpage
    % \subsection*{Evaluation considerations}

    \subsection*{Algorithm Evaluation}
    
    As the algorithm's task was to place an individual detection anywhere on a signal, a problem presents itself when comparing detections to a ground truth.
    If the algorithm produces a false positive, splits a single breath into two or more breaths, or any case in which an extra detection is inserted, a misalignment between the list of detections and annotations is created where the detections placed after the false positive have an index that corresponds to the index of a later annotation than it should. The error compounds after each false positive.
    % 
    % \begin{figure}[h!]
    %     \centering
    %     \includegraphics[width=0.8\linewidth]{Graphics/Diagrams/FalsePositiveError.png}
    %     \caption{Misalignment of manual annotations and detections created due to false positive}
    %     \label{fig:false_positive_misalignment}
    % \end{figure}
    % 
    The same misalignment error is created when the algorithm produces a false negative, except the misalignment is now reversed, i.e., each detection after the false negative has an index corresponding to the index of an earlier annotation than it should. As with the previous case, the misalignment error compounds after each false negative. 
    
    % \begin{figure}[h!]
    %     \centering
    %     \includegraphics[width=0.65\linewidth]{Graphics/Diagrams/FalseNegativeError.png}
    %     \caption{Misalignment of manual annotations and detections created due to false negative}
    %     \label{fig:false_negative_misalignment}
    % \end{figure}
    
    Due to the possibility of misalignment errors, it was not possible to naively compare the list of detections and annotations, and an extra step had to be performed to match detections to their corresponding annotations.
    This alignment problem was solved using a matrix containing the percentage overlap of all available pairs of detections and annotations calculated using \eqref{eq:overlap_percentage}. 
    This matrix is referred to as the overlap matrix and simplifies the process of finding which detection corresponds to which annotation, whether or not a given detection is a false positive or not, and whether a given annotation corresponds to a detection or is a false negative. 
    % A simulated overlap matrix is displayed in \figRef{fig:overlap_matrix}.
    % 
    Given an overlap matrix $A$ of a list of detections $X$ and a list of annotations $Y$, the overlap of any $X[i]$ and $Y[j]$ can be accessed in $A[i,j]$. 
    % This problem is solved using an overlap matrix, the matrix is constructed by calculating the percentage overlap between every pair of detection and label using \eqRef{eq:overlap_percentage}. A simulated overlap matrix is visualised in \figRef{fig:overlap_matrix}.
    % The weighted overlap value is calculated like so: 
    % \begin{equation}
    %     O_{w}(A, B) = \frac{2|A {\displaystyle \cap} B|}{|A|+|B|} 
    % \end{equation}
    % This captures the overlap value between two time spans as a percentage of the cumulative span length. This metric prevents the algorithm from either making few massive detection that cover all labels, or making multiple smaller detection, each covering a minuscule portion of the label. The percentage overlap calculation is also symmetric, meaning that the calculation will produce the same value for any pair of intervals. 
    
    % \begin{figure}[h!]
    %     \centering
    %     \includegraphics[width=0.7\linewidth]{Graphics/overlap_drawn.png}
    %     \caption{Example overlap matrix}
    %     \label{fig:overlap_matrix}
    % \end{figure}
    
    Using an overlap matrix, a detection corresponding to any annotation could be found by locating the index of the maximum overlap value in the overlap matrix column for that annotation. 
    To be counted as a correct detection for a given annotation, a detection must have a weighted overlap value of over 80\% with that annotation.
    If an annotation had no value above that threshold in its column in the overlap matrix, the breath was counted as having been missed by the algorithm (false negative).
    Similarly, if a detection had no value above the threshold in its row in the overlap matrix, the detection was counted as a false positive.
    Due to the restriction that detections may not overlap, it was impossible for two detections to correspond to the same annotation.
    % 
    % The simulated overlap matrix in \figRef{fig:overlap_matrix} includes an example of a false positive in row 3 which describes the fourth detection. The fourth detection has a maximum percentage overlap value in column 3, however, the overlap is below the threshold, so it is counted as a false positive. 
    % 
    This paper uses the precision (ratio of true positives to true positives and false positives), recall (ratio of true positives to true positives and false negatives), and F1 score metrics to estimate the accuracy of the algorithm.
    
    In addition to the precision, recall, and F1, additional statistics were collected on the placement error of the detections that were counted as correct. Those include the length of detections vs. the length of the annotations.
    The start and end error was calculated using the following 2 formulae:
    
    \begin{equation}
        \text{start error} = s - \hat{s} \qquad \text{end error} = e - \hat{e}
        \label{eq:start_and_end_error}
    \end{equation}
    
    Where $s$ is the annotation start, $e$ is the annotated breath end, $\hat{s}$ is the predicted breath start, and $\hat{e}$ is the predicted breath end. 
    
    % \begin{figure}[h!]
    %     \centering
    %     \includegraphics[width=0.7\linewidth]{Graphics/Diagrams/breathfinder-ErrorExplanation.png}
    %     \caption{Visualisation of detection error calculation}
    %     \label{fig:error_visualisation}
    % \end{figure}
    
    The algorithm has 4 main parameters; the analysis window length, the overlap threshold, the correlation cut-off for the sine fitting procedure, and the probability threshold for the  filtering process. To gauge the effect these variables have on the algorithms performance, the evaluation was repeated for a range of values for each variable.

\section*{Results}

    The algorithm was evaluated on two sets of manually annotated intervals, the first set containing a relatively high amount of SDB events, and the second set being sampled from a population of 25 participants. The results of the performance evaluation are summarised in \tabRef{tab:performance}, and the placement errors are shown in \tabRef{tab:length_errors}
    
    \begin{table}[H]
        \caption{Evaluation results of the RCI algorithm }
        \begin{tabularx}{\linewidth}{Y | Y Y Y}
            \hline
            Evaluation subset & Mean precision & Mean recall & Mean F1 \\ \toprule
            A & 0.94              & 0.94           & 0.94       \\
            B & 0.93              & 0.95           & 0.94   \\  \bottomrule 
        \end{tabularx}
        \label{tab:performance}
    \end{table}
    % The algorithm achieved 0.94 F1 score for both evaluation subsets,  
    
    The algorithm achieved on average 0.94 precision for \ESA and 0.93 for \ESB, this means that only 6\%, and 7\% of detections were classified as false positives for \ESA and \ESB respectively. The recall for \ESA and \ESB was 0.94 and 0.95 respectively, meaning that the algorithm only missed 6\% of breaths in \ESA and 5\% of breaths in \ESB.
    Two intervals in \ESB had noticeably worse results, with F1 scores of 0.79 and 0.81. Upon visual inspection the errors seemed mainly to be due to incorrect manual annotations, and noise in the signal during those intervals. 
    If these 2 intervals were omitted, the mean F1 of the algorithm increased to 0.95 for \ESB.
    The algorithm's performed noticeably worse for one interval in \ESA than for the others, its precision being 0.76, recall being 0.963, making for an F1 score of 0.854. This was due to the interval being relatively short, and the beginning of the signal being dominated by a movement event, causing the algorithm to misclassify the movement as breaths. 
    On the other hand, the algorithm achieved perfect precision for two intervals in \ESA and one in \ESB, all of which contained no SDB events, and only stable breathing.
    The recall was slightly more stable than the precision for both sets, with the standard deviation being 0.054 for the recall, and 0.058 for the precision. 

    % For \ESA, the algorithm achieved on average 0.94 precision, and 0.94  recall. This means that of the total number of breaths detected, 5.8\% were false positives, and the algorithm missed 5.2\% of the labelled breaths.
    % In line with the results from \ESA, the algorithm achieved on average 0.93 precision for \ESB, meaning that of the total number of detections, only 6.4\% were false positives. The consistent results over both \ESA and \ESB indicate that the choice of subject is not a significant factor in terms of algorithm performance, although larger datasets are needed to confirm this.
    % % 
    % While this difference is less than for \ESA, this further suggests that the algorithm has a greater tendency to produce false positives rather than false negatives. 
    % % TODO: Move to discussion
    % % suggesting that the presence of movement in the beginning of the signal does not impact later detection. 
    % % 

    \begin{table}[H]
        \centering
        \caption{Placement errors of the RCI algorithm.}
        \label{tab:length_errors}        
        \begin{tabularx}{\linewidth}{Y | Y Y Y Y}
            \hline 
            Evaluation subset & Annotation mean breath length & Detection mean breath length & Mean abs. start error (seconds) & Mean abs. end error (seconds)  \\ \toprule
            A & 2.57    &    2.76    &    0.16    &    0.24   \\ 
            B. & 3.56  &  3.88  &  0.23  &  0.30 \\ \bottomrule%\hline
        \end{tabularx}
    \end{table}

    The mean start error for both \ESA and \ESB was approximately 6.4\% of the mean breath length.  
    The mean end error for both sets was greater, being 10\% and 8.4\% of the mean breath length for \ESA and \ESB respectively. 
    When visually inspected, the alignment of the detections and the thoracic RIP signal was very high for both \ESA and B.

    % The start and end errors generally fell within 6-10\%, with the mean absolute start error only being 6.25\% of the average breath length. The absolute end error was more significant, or 0.24 seconds, being at about 10\% of the actual mean breath length, however, when visually inspected, the error was not visibly discernible.
    
    % Similarly to the results from \ESA, the absolute start error for \ESB was on average 0.23 seconds, or 6.4\% of the mean breath length, and similarly to \ESB, the absolute end error was greater, or about 8.4\% of the average breath length. Upon visual inspection of the detection placement, the start and end errors were not visually discernible.
    % % \input{Tables/ESA_lengths}
    % % 
    
      \subsection*{Sensitivity Analysis Results}
  
    \newcommand{\ptwidth}[0]{1\linewidth}

    The analysis window length was the length in seconds of the window that the algorithm searches for breaths in at each step. 
    \begin{figure}[h!]
        \centering
        \includegraphics[width=\ptwidth]{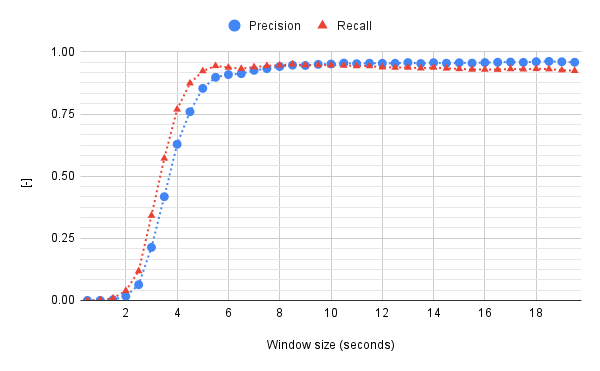}
        \caption{Algorithm Recall and Precision sensitivity to window length.}
        \label{fig:window_length_sensitivity}
    \end{figure}
    The results of the sensitivity estimation can be seen in  \figRef{fig:window_length_sensitivity}, and shows that both precision and recall rise sharply as the window length reaches approximately 6 seconds, and plateaus at approximately 8 seconds. The reason for the sharp rise in performance between 2-6 seconds is most likely that the window can not reliably fit two cycles of the respiratory cycle until the window becomes longer than twice the average length of breath in the dataset.
    
\begin{figure}[h!]
        \centering
        \includegraphics[width=\ptwidth]{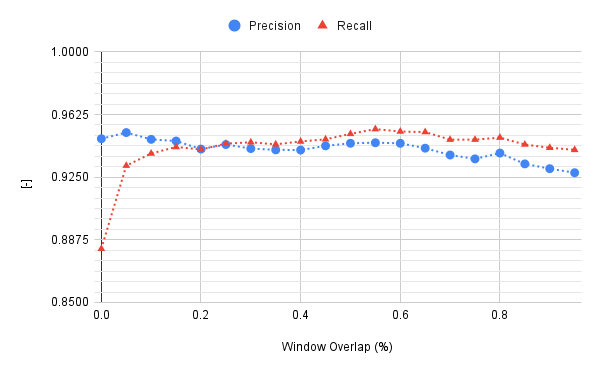}
        \caption{Algorithm Recall and Precision sensitivity to overlap percentage.}
        \label{fig:overlap_percentage_sensitivity}
    \end{figure}

    %  The overlap, which determines how much of the previous window is included in the next window, was chosen to be 95\% during the development of the algorithm.
    The overlap percentage is the amount of the previous window included in the next window as the analysis window advances.
    The effect of this parameter is shown in \figRef{fig:overlap_percentage_sensitivity}, which suggests that the algorithm performs noticeably worse only in terms of recall when the overlap percentage is around 0. This can be explained as the algorithm may miss breaths as the window skips either entirely, or partly over them. The precision seems largely unaffected, which indicates that the number of false positives drops proportionally with the number of true positives.

\begin{figure}[h!]
        \centering
        \includegraphics[width=\ptwidth]{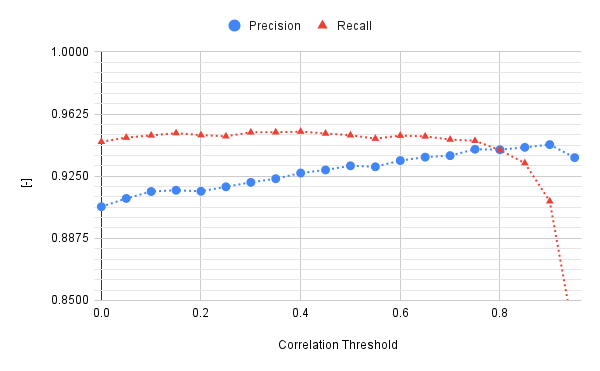}
        \caption{Algorithm recall and precision sensitivity to correlation threshold.}
        \label{fig:correlation_threshold_sensitivity}
    \end{figure}

    The correlation threshold dictates how much a breath candidate must resemble a model breath in terms of Pearson correlation (\figRef{fig:correlation_threshold_sensitivity}). It is meant to filter out waveforms that may only superficially resemble breaths, however, still forming peaks in the sine-correlation function. 
    As the correlation threshold increases, the precision improves. This can be interpreted as the criterion for what "looks like a breath" becoming stricter, thus eliminating more false negatives.
    The recall seems unaffected by this criterion until the threshold reaches approximately 0.8, at which point it sharply drops. This is to be expected, since the template waveform is only an estimation of the general shape of a breath in the signal, and thus the correlation with the signal is not expected to be perfect.

    The probability threshold parameter is used to discard breaths that are considered too unlikely. To estimate the effect of this variable on the performance of the algorithm, the evaluation was repeated for a range of probability thresholds. 
    \begin{figure}[h!]
        \centering
        \includegraphics[width=\ptwidth]{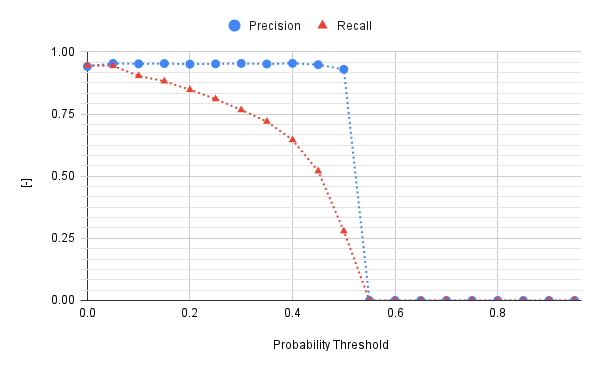}
        \caption{Algorithm Recall and Precision sensitivity to probability threshold.}
        \label{fig:pthresh_relation}
    \end{figure}
    As \figRef{fig:pthresh_relation} indicates, the absence of this filtering step seems to have little effect. The precision is least affected by the probability threshold, while the recall drops sharply off as the threshold increases. This is reasonable, since as the threshold increases more legitimate breaths are discarded, thus negatively affecting the recall until the threshold reaches approximately 0.55, at which point all breaths are discarded. The stability of the precision suggests that the rate of false positives drops proportionally to the rate of true positives as this parameter approaches 0.5. 
    The reason for the falloff of both the precision and recall at 0.5 is that the maximum possible value of the probability estimator is 0.5, any value above 0.5 will therefore cause the filtering process to discard all detections.

\section*{Discussion}

% The Discussion should be succinct and must not contain subheadings.
This paper proposes a novel algorithm designed to perform RCI on the thoracic RIP signal, based on signal processing and statistical methods.
The algorithm achieved an F1 score of 0.94 when detecting breaths during sleep during both SDB events and over multiple participants, meaning that the algorithm detects around 94\% of breaths correctly, with 6\% false negatives. Of the detections made by the algorithm, approximately 95\% are correctly placed breaths, with only 5\% being false positives. 

% Accuracy in terms of different approaches
The algorithm achieved accuracy either superior \cite{Rosenwein2014-nz}, or comparable \cite{LopezMeyer_2011, Hult2000-lx} 
to prior work.
A notable problem with comparison to prior work is that there exists no standardised method of evaluating RCI algorithms, and thus different works approach the evaluation differently, making comparison to that work difficult, if not at times impossible \cite{Palaniappan2017-jv, wlodarczak2019respinpeace, 95756}. 
% 
% The effect of snoring has not fully been examined on the performance of the algorithm, however, due to the scale agnosticism of the correlation used in the sine-fitting procedure, any flow limitation caused by snoring is unlikely to affect the performance greatly.
% 
Currently, the algorithm is only evaluated on RIP signals collected with a 25$\textit{Hz}$ sampling frequency. The algorithm is designed to be independent of the sampling frequency of the signal, but requires a similarly rigorous evaluation at other sampling frequencies. 
In this paper, the algorithm is validated on intervals containing significant movement, respiratory events, as well as various sleep stages.
% 

% This work also sets an example for evaluation of RCI tasks.
% 
The evaluation found that the detection rate was not meaningfully influenced by respiratory events or physiology, however, the most impactful factor in terms of detection rate seemed to be artefacts. On the other hand, artefacts such as movement or signal failure only cause the algorithm to produce errors where the artefacts occur, and cause no errors for future detections, indicating that the algorithm can easily recover from a noisy period.
The detection error of the correctly detected breaths is expressed in the mean absolute start and end error, the algorithm tends to produce greater end errors than start errors. The mean end error, however, was less than 8\% of a mean breath length, and upon visual inspection, was not discernible to the human eye. 
The start and end errors of both sets may be partially explained by the fact that the manual annotations did not observe the restriction that only one breath can take place at any moment imposed by this work's definition of the respiratory cycle, effectively introducing small sections of the annotations that at most one detection can overlap with thus negatively impacting the metrics artificially.

% future potential
Despite the fact that the algorithm is originally designed for use in sleeping individuals, it could be used to research respiration during speech, exercise, emotional response analysis, and other applications, provided that the proper evaluation of the output correctness is performed.
The total number of participants used for the evaluation of the algorithm was 25 individuals. In contrast to the literature, this number is comparable, as the range of the number of individuals used for testing similar tasks ranges between none reported, 4, 75, and 140 participants
\cite{95756, wlodarczak2019respinpeace, LopezMeyer_2011, Rosenwein2014-nz, Palaniappan2017-jv}. 
In future work, the algorithm should be evaluated on a much larger dataset.
The algorithm is designed to work on the thoracic RIP signal, but in theory should also work on the abdominal RIP signal. However, this requires validation to assess the validity of the results.
%

% Final word
Due to the high detection rate of the algorithm and the relatively low rate of false positives, the authors suggest that the proposed algorithm can be reliably used for future research into the nature of respiration during sleep based on RCI-based adaptive segmentation. However assessment on larger datasets is needed to evaluate the performance of the algorithm when faced with a more diverse range of respiratory events such as central apneas.

\section*{Public Availability}
    The algorithm described in this work has been implemented in Python, and made open-source under the GNU licence. It is available both via GitHub: \href{https://github.com/benedikthth/BreathFinder}{github.com/benedikthth/BreathFinder}, and the Python pakage manager:
    \href{https://pypi.org/project/BreathFinder/}{pypi.org/project/BreathFinder/}.

\printbibliography

\clearpage

\newcommand{\bibimage}[1]{\includegraphics[width=1in,height=1.25in,clip,keepaspectratio]{#1}}

\begin{IEEEbiography}[{\bibimage{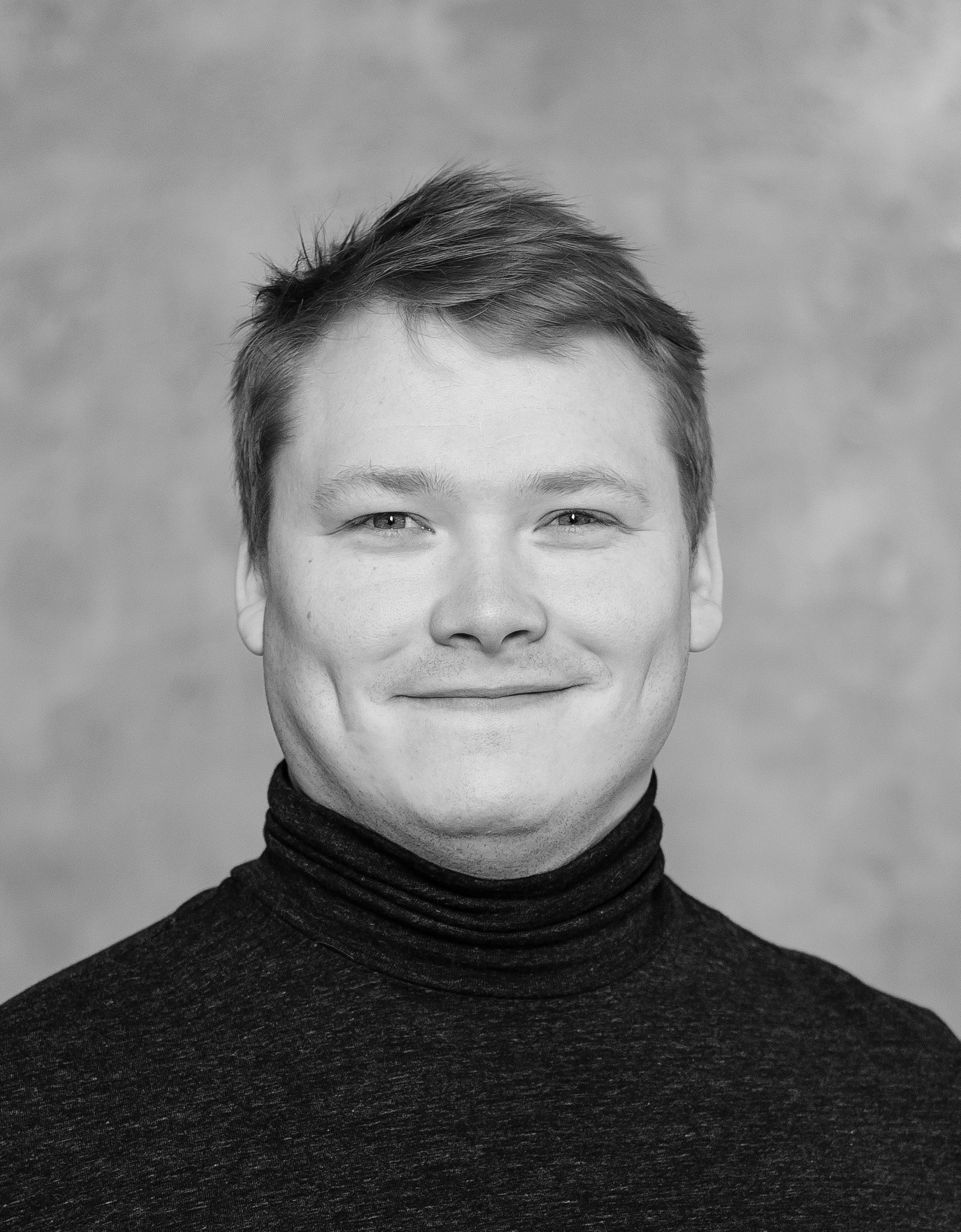}}]{Benedikt Holm} 
    is a Ph.D. student in Reykjavik University working with the Sleep Revolution. He received his M.Sc. degree in computer science from Reykjavik University, Reykjavik, 2021, where he worked on using machine learning to identify obstructive apnea in individual respiratory cycles. His research interests are focused on using adaptive segmentation and data science to optimise automation in sleep research.
\end{IEEEbiography}

% {\includegraphics[width=1in,height=1.25in,clip,keepaspectratio]{a3.png}}

\begin{IEEEbiography}[{ \bibimage{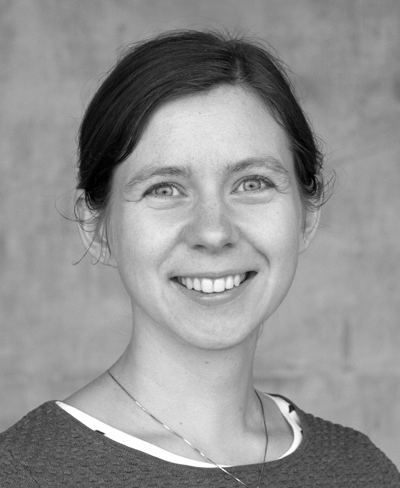} }]{Maria Óskarsdóttir} 
    %https://svefnsetrid.ru.is/member/?smid=552
    is an Assistant Professor at the Department of Computer Science at Reykjavík University.
    She holds a PhD in business analytics from KU Leuven, Belgium, and a master's degree in Mathematics from the Leibniz University Hannover, Germany. Her research is focused on practical applications of data science and analytics whereby she leverages advanced machine learning techniques, network science and various sources of data with the goal of increasing the impact of the analytics process and facilitating better usage of data science for decision making, currently focusing on sleep measurements. 
    % María is the director of the Master programs in Data Science and Applied Data Science at Reykjavík University. She is currently supervising two PhD students working on data-driven health.
\end{IEEEbiography}

% \vfill

\begin{IEEEbiography}[{ \bibimage{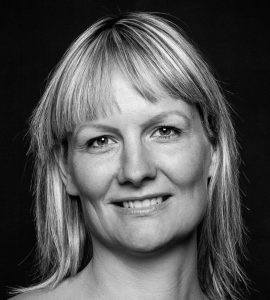} }]{Erna Sif Arnardóttir} 
    is an Assistant Professor at the Departments of Engineering, and Computer Science at Reykjavik University and the Director of the Reykjavik University Sleep Institute (RUSI).
    She also has an advisory position at the Landspitali – The National University Hospital of Iceland. She has over 15 years of experience within the field of clinical and scientific sleep research. 
    % Currently, she is the President of the Icelandic Sleep Research Society and a board member of the European Sleep Research Society (ESRS) task force “Beyond the apnoea-hypopnea index – Standardization of additional measures in adult sleep studies for obstructive sleep disordered breathing”, which just published its first paper (Pevernagie et al 2020).
    % 
    % She also coordinated the launch of the ESRS Early Career Research Network. She has authored over 40 published peer-reviewed papers and one book chapter. Currently, Dr. Arnardóttir and her team are focusing on new ways to assess sleep disordered breathing severity, from habitual snoring to severe OSA. 
    % 
    Erna is the Principal Investigator in a grant from EU Horizon 2020, from 2021 to 2025. The project name is SLEEP REVOLUTION, is an inter-disciplinary and international research and development project with 37 partner institutions and companies in Europe and Australia. aim is to bring personalised medicine to sleep apnoea research, which aligns completely with the research aims of this project.
    Dr. Arnardóttir is the Icelandic Principal Investigator (PI) of NordSleep, a consortium project between Iceland, Finland and Norway. Her Google scholar profile shows >2000 citations, an H-index of 23 and i10-index of 32. 
% At present she supervises 1 postdoctoral researcher, 3 PhD students, 6 M.Sc students and 1 B.Sc student in the field of sleep research.
\end{IEEEbiography}

\begin{IEEEbiography}[{ \bibimage{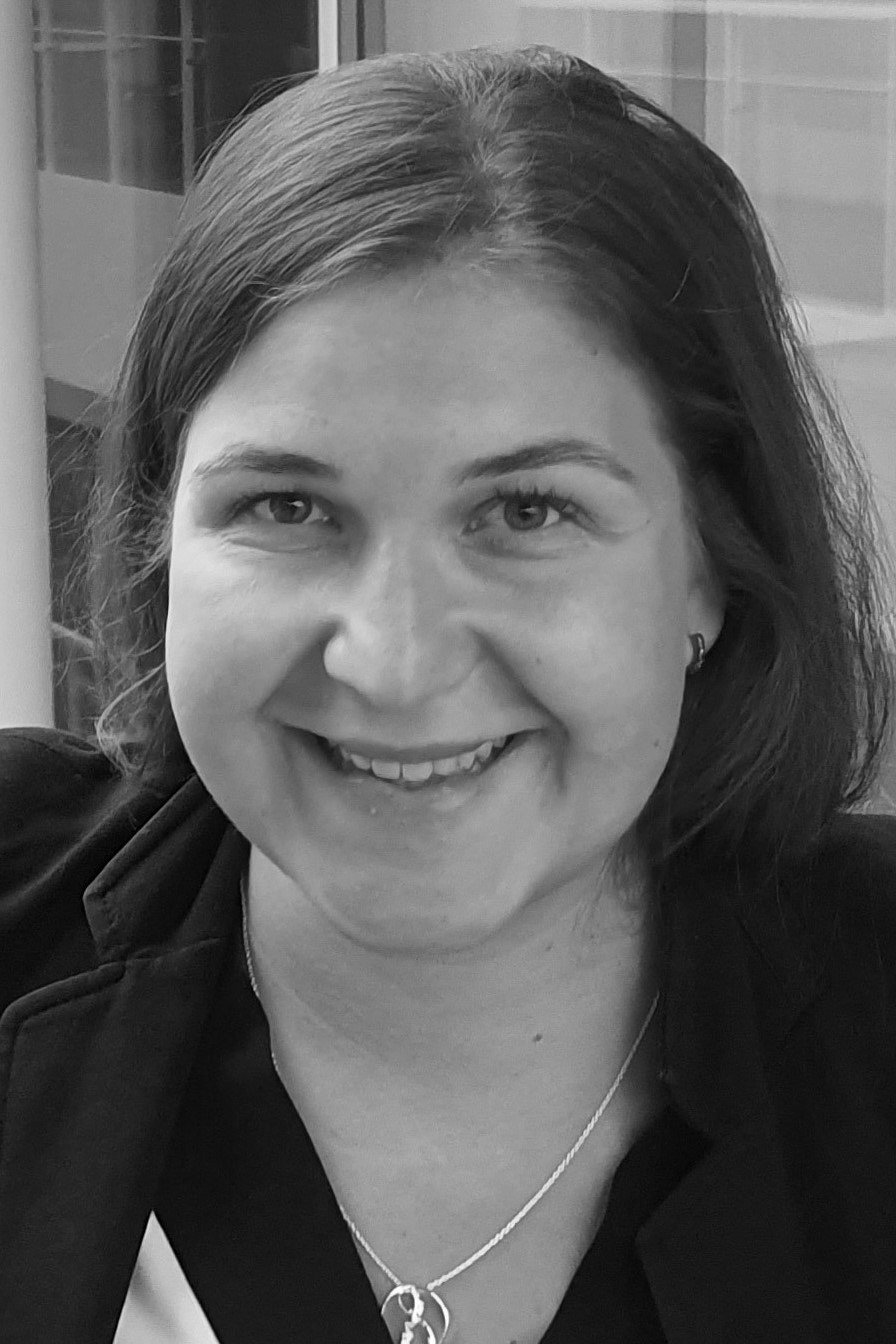} }]{Marta Serwatko}
    received her B.Sc. and M.Sc. degrees in biomedical engineering from Reykjavik University in 2013 and 2016, respectively. She has been working in sleep research since 2013 and her main research interests include sleep research and investigation of how and to what degree the disturbed sleep, due to decreased respiration, affects both adults and children, and further, what are the health consequences.
\end{IEEEbiography}

% \vfill

\begin{IEEEbiography}[{ \bibimage{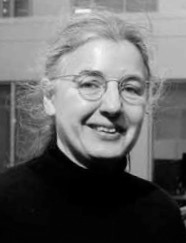} }]{Jacqueline Clare Mallett} 
    is an Assistant Professor in RU’s Department of Computer Science.
    She is a multi-disciplinary computer scientist with a PhD in computer science from MIT and 30 years of industry and research experience in designing, building, and troubleshooting distributed systems, including real-time critical systems, high performance computing, signal processing, wide-area networking and computer security. She has worked on secure advanced server infrastructure for high frequency trading and consumer appliances, and teaches Computer Security at both University of Iceland and RU.
\end{IEEEbiography}

\begin{IEEEbiography}[{ \bibimage{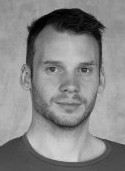} }]{Michal Borsky} 
    is a research specialist at the RU Department of Computer Sciences.
    Dr. Michal Borsky has held an interest in signal processing since his master’s studies. This interest made him focus on speech recognition for his PhD degree, which he obtained at Czech Technical University in Prague. Since then, he has worked at Reykjavik University, focusing on pathological speech, and later on sleep disordered breathing and snoring. His current interests include artificial intelligence, information theory, and human biology.
\end{IEEEbiography}

% \vfill

\end{document}